\documentclass{aipproc}
\input{epsf} 

\layoutstyle{6x9}
 
\begin{document}
\title{Phenomenological implications of brane world 
scenarios with low
tension\footnote{Talk given by A. Dobado in the X 
Mexican School of
Particles and Fields, Playa del Carmen, M\'exico, 2002}}

\author{J.A.R. Cembranos, A. Dobado, 
A. L. Maroto}{
address={Depto. de F\'{\i}sica Te\'orica,
Universidad Complutense de Madrid, 28040 Madrid, Spain}}

\begin{abstract}
The recent proposal of theories with compactified large 
extra dimensions is
reviewed. We pay especial attention to brane world 
models with low tension
where the only relevant degrees of freedom at low energies 
are the Standard
Model (SM) particles and the branons, which are 
transversal brane oscillations.
By using an effective Lagrangian, we study some 
phenomenological
consequences of these scenarios in a model 
independent way. 
\end{abstract}
\maketitle

\section{Are extra dimensions real  or are they  
just science fiction?}

Most of the motivations for considering extra dimension have a theoretical
origin. In the last thirty years most of the new developments in theoretical
high energy physics required the introduction of extra dimensions. Well known 
examples are modern Kaluza-Klein theories (KK), where the isometries of the
extra dimensions appear as gauge symmetries, Supersymmetry (SUSY), which in the
superfield formulation can be understood as a symmetry involving extra
Grassmann dimensions, Supergravity (SUGRA), Superstrings defined
consistently only in 10 dimensions and M-theory, which is supposed to be the
eleven dimensional theory underlying the five known superstring theories and
eleven dimensional SUGRA. One important exception to this rule are the Grand
Unified Theories (GUT) but still the most interesting examples from the
phenomenological point of view are  supersymmetric since they give rise to
gauge coupling unification. 

The first attempts to extend general relativity to include 
electromagnetism
date back to Theodor Kaluza and Oscar Klein \cite{KK}. The
discovery of 11D SUGRA produced a revival of the KK ideas in the 
early 80's.
The first string revolution, triggered by the realization that 
superstrings can
provide a theory of gravitation instead of a theory of strong interactions,
the  cancelation of the anomalies and the discovery of the heterotic 
string
which opened the door for phenomenology, traslated the interest to $10$
dimensions with six dimensional compactified spaces (Calabi-Yau,
orbifolds...). The second string revolution of the 90's introduced
new ideas such as non-perturbative strings, dualities, branes and the
unification of the five known superstring theories in the conjectured
M-theory (these two periods of development of  
string theory can be reviewed in \cite{GSW} and \cite{P} respectively). 

The main phenomenological problem of the old string theories is that they
could not be tested since stringy effects were expected to appear at the Plank
scale $M_P$ which is of the order of $10^{19}$GeV. However the new ideas
coming from M-theory have inspired new scenarios that could be testable. These
scenarios were developed in principle to address the hierarchy problem 
by putting 
it in a different setting. The first one was proposed by Arkani-Hamed,
Dimopoulos and Dvali (ADD) \cite{ADD}. The main idea is that our universe is a
3-brane living in a higher $D=4+N$ dimensional space (the bulk space) being
the extra dimensions compactified to some {\it small} volume (Brane World
scenario). In this picture the SM particles are confined to the 3-brane but
gravitons can propagate along the whole bulk space. Now the fundamental scale
of gravity is not the Planck scale any more but another scale $M_D$ 
which is
supposed to be of the order of the electroweak scale in order to solve 
the
hierarchy problem. Then it is possible to find the relation 
$M_P^ 2=V_N M_D^
{2+N}$, where $V_N$ is the volume of the compatified extra dimension manifold
$B$. Thus the hierarchy between the Planck and the electroweak (TeV) 
scale
is generated by the {\it large} volume of the extra dimensions. The typical
size $R$ of the extra dimensions ranges from a fraction of mm for $N=2$ 
to
about 10 fm for $N=6$ (the case $N=1$ is already ruled out by the
observations in our solar system). The most interesting property of the  ADD
scenario (in which we will concentrate in the rest of this work) is that it is
compatible with the present experimental data, but at the same time it 
gives
rise to many new phenomena that could be tested in the near future.

There are also scenarios where the scale of the extra dimensions 
$R$ is of
the order of 1 TeV$^{-1}$  \cite{A}. All  or some of the SM particles are
allowed to propagate along the bulk. This set up is quite appropriate for model
building and to deal with gauge coupling unification, SUSY breaking, 
neutrino
spectrum, fermion masses and many other things even if it does 
not solve the
hierarchy problem. In addition there are also  models where the hierarchy is
generated by the curvature of the extra dimensions as it is the case of the
Randall-Sundrum model where the geometry of the space-time is $AdS_5$ and thus
it cannot be factorized \cite{RS}.

\section{Is our Universe a 3-brane?}

The most obvious consequence of the the Brane World scenario is the
modification of the Newton's Law at short distances i.e. distances of 
the
order of $R$. Some recent experiments are trying to test the Newton 
Law's at
the submillimeter scale to explore this possibility (see for instance
\cite{W}).
From the point of view of particle physics the main new effects in the ADD
scenario are related with the KK mode expansion of the bulk gravitational field
\begin{equation}
g_{\mu\nu}(x,\vec y)=\sum_{\vec{k}}
g_{\mu\nu}^{\vec k}(x)
e^{i  \vec k . \vec y/ R} 
\label{KKmode} \end{equation}
where a toroidal compactification has been assumed for simplicity, $\vec y$
represents the $N$ extra dimension coordinates and $\vec k$ is a $N$
dimensional vector with components $k^a=0,1,2,...$. 
Therefore a bulk graviton
can be understood as a KK tower of four dimensional massive gravitons 
with
masses of the order of $k/R$ with $k$ being any natural number (for the
$N=1$ case) so that the distance in the mass spectrum 
between two consecutive KK gravitons is 
of the order of $1/R$. This means in particular that the KK graviton 
spectrum
can be considered as approximately continuous for large extra 
dimensions. In
principle we expect two kinds of effects from the KK graviton tower, namely
graviton production and virtual effects on other particle production or
observables (see for instance \cite{HS} and references therein). The rates for
the different processes can be computed by linearizing the bulk gravitational
field and by coupling the graviton field to the SM energy momentum tensor
$T_{SM}^{\mu\nu}$. Then expanding the gravitational field in terms of the KK
modes one finds the corresponding Feynman rules. As an example one could
consider the process $e^+e^- \rightarrow \gamma+ G$ where $G$ stands for any
KK graviton. To compute the total cross section one must sum (integrate for
large extra dimensions) over all the KK gravitons. The result is
\begin{equation} \sigma_{KK}  \simeq \frac{1}{M_P^2}(\sqrt{s}R)^ N  \simeq
\frac{1}{M_D^2} \left( \frac{\sqrt{s}}{M_D} \right)^ N  \label{skk} 
\end{equation}
Note that the total cross section is suppressed by powers of 
$M_D$ which is
supposed to be of the order of 1 TeV instead of powers of $M_P$ which would
make the cross section completely negligible. The signature of these events
would be one single photon plus missing energy with a continuous spectrum.
Virtual effects can be taken into account by considering the KK tower
propagator \begin{equation}
\sum_{\vec{k}}\frac{1}{-p^2+ k^2/R^2}
\label{KKpropagator} 
\end{equation}
This propagator produce divergencies for more than one extra dimension
even at the tree level that require regularization by introducing some
ultraviolet cutoff. This is one of facts that has given rise to the development
of the so called deconstructing \cite{ACG} or aliphatic \cite{CHPW} idea where
the extra dimensions are {\it latticed}.

\section{Brane fluctuations}

However there is a more physical way to deal with the above problem. So far we
have asssumed that the world brane is completely rigid, i.e. it has infinite
tension. But rigid objects do not exist in relativistic theories. When brane
fluctuations are taken into account two new effects appear. First of all we
have to introduce new fields which represent the position of the brane
in the bulk space $(x^{\mu},y^a=\pi^a(x))$. These fields are the Goldstone
bosons (GB) corresponding to the spontaneous symmetry breaking (SSB) of the
translational invariance produced by the presence of the brane (branons). When
the effect of the branons is taken into account it is possible to show that
the coupling of the SM particles on the brane to any bulk field is
exponentially suppressed by a factor 
$\exp(-M_{KK}^2M_D^2/(8\pi^2f^4))$, 
where
$M_{KK}$ is the KK mode mass and $\tau=f^4$ is the brane tension \cite{GB}.
This effect  solves the above mentioned divergency problem. The other
important consequence of this effect is  that in the case where 
$f \ll M_D$
i.e., when the brane tension scale is much smaller than the fundamental scale
of gravity (in the $D$ dimensional space), the KK modes decouple from the SM
particles on the brane. The conclusion is that for flexible enough branes the
only relevant degrees of freedom at low energies in the ADD scenario are the SM
particles and the branons. As other GB, branons are expected to be nearly
massless \footnote{Some branon mass $M$ could be expected from explicit symmetry
breaking effects as it happens with  pions, which are the GB related to the
 SSB of the chiral symmetry of strong interactions} and weakly
interacting at low energies, and their interactions can be described by an
effective lagrangian. For homogeneous extra dimension compact 
space $B$, this
effective lagragian has the structure of a non-linear sigma model where the 
coset space is just (is isomorphic to) the space $B$ \cite{DoMa}. In addition
the branon couple by pairs to the energy-momentum  tensor of the SM, 
$T_{SM}^{\mu\nu}$ through the tensor $\partial_{\mu}\pi^ a\partial_{\nu }\pi^a$
quite in the same way  as  gravitons do. Thus it is possible to
obtain the different Feynman rules and amplitudes for branon production
starting from SM particles in terms of the SM parameters, the brane
tension scale $f$, the number of branons $n$ ($n=N$ for homogeneous spaces $B$)
and the possible branon mass $M$ (\cite{ACDM}).
The typical signature for branon production would be missing energy with
continuous spectrum, but in this case the missing energy spectrum is 
continuous
due to the fact that branons are produced in pairs. In addition 
by studying
branon production one is probing the brane tension scale $f$ 
instead of the
fundamental gravitation scale $M_D$ which is probed 
in graviton production. In
the following we will consider the branon phenomenology in the 
ADD scenario
with a flexible brane $f \ll M_D$  in order to set some bounds on the brane
tension and the branon mass from particle physics, astrophysics and cosmology.

\section{Bounds from Z and W decays}

 In the Standard Model, the full decay width of the Z
 boson has three types of contributions, coming from charged leptons,
 hadrons
 and neutrinos respectively. The third one corresponds to the
 so called invisible width, since neutrinos escape without producing
 any signal in the detectors. In principle, 
 light branons ($M<M_Z/2$)
 being stable and chargeless
 particles, could also give rise to invisible decays
 of Z and therefore
 to deviations with respect to the SM predictions.

 The precision measurements done mostly at LEP-I set stringent
 limits on the Z invisible decay width, which
 can be translated into strong bounds 
 on the brane tension $f$ and the
 branon mass. Thus, at the 95$\%$
 confidence level, the
 variation of the Z invisible width due
 to decays into branons $\Gamma_Z^{b}(n,f,M)$ cannot be 
 larger than 2.0 MeV
 \cite{LEP}.

 We use the results from \cite{ACDM} for the 
 first order contribution from
 branons to this width, which is given by the decay of Z into two
 branons and two neutrinos:
 $Z\rightarrow\bar\nu\,\nu\,\pi\,\pi$.
 Imposing the LEP-I limit on this decay width
 we get exclusion plots in the $f-M$ plane (see Figure 1, left
 for models with a single branon).
 For $n$ massless branons,
 the bound on $f$ reads:
 $f\,>\, 11.9\, n^{1/8}\, \mbox{GeV}$.

 Concerning the $W^\pm$ decay, 
 there are mainly two new channels contributing to this process.
 First, the decay of $W^\pm$ into two branons
 and two leptons:
 $W^-\rightarrow l^-\,\bar\nu\,\pi\,\pi$.
 Such leptons can be an electron and an electron antineutrino for
 the $W^-$ decay, their antiparticles for $W^+$, or the analogue
 pairs of leptons in the rest of families. 
 Second, we have the $W^\pm$ decay into
 two branons and a quark-antiquark pair:
 $W\rightarrow q\,\bar q\,\pi\,\pi$.
 In principle there should be also a channel including gluons
 in the final state, however in the SM, such channel is
 suppressed by a coefficient  $\alpha_S(M_W)/\pi=0.04$ and
 for that reason we will also ignore it
 in the present case. In any case, the result we will obtain
 will be  an strict lower bound to the true decay width into branons.

 The bounds on the contribution from new physics to the
 $W^\pm$ width $\Gamma_W^{b}< 240$ MeV, are weaker than the Z
 case. 
 This limit translates into the following one for massless branons:
 $f\,>\, 6.9\, n^{1/8}\, \mbox{GeV}$.
 For a single branon, we have plotted the exclusion region
 for the variables $f$ and $M$ in Figure 1 (left) \cite{ACDM}.

 \section{Bounds from direct searches}

 The differential cross section of the process:
 $e^+\,e^-\rightarrow\pi\,\pi\,\gamma$
 including the two photon polarizations reads \cite{ACDM}:
 \begin{eqnarray}
 \frac{d\sigma_\gamma}{dx\;d(\cos\theta)}&=&\alpha_{EM}^2\frac{2s
 (s(1-x)-4M^2)^2 n}{61440f^8\pi^2}\sqrt{1-\frac{4M^2}{s(1-x)}}
 \nonumber\\
 &&\left[x(3-3x+2x^2)-x^3\sin^2\theta
 +\frac{2(1-x)(1+(1-x)^2)}{x\sin^2\theta}\right]\nonumber \\
 \end{eqnarray}

 Such expression could be used in
 direct searches of branons in colliders. However, we
 have calculated the contribution to the
 total cross section of processes involving a single
 photon in the final state. 
 The total cross section can be calculated analytically, although
 we will not show it here since the final
 expression is quite long. However the result for massless 
 branons (M=0) is much simpler:
 \begin{eqnarray}
 \sigma_\gamma(M=0)&=&\frac{n A s^3}{f^8}.
 \label{zeromass}
 \end{eqnarray}
 where the  constant depends on the detection limits.
 For the LEP-II experiments  we get $A=4.54\times 10^{-7}$.
 This has allowed us to obtain new
 bounds on $f$ assuming no observation at LEP-II:
 $f \geq 138 n^{1/8} \mbox{GeV}$. 
 
 The results in the massive case for one  branon assuming no 
 observation is plotted in Figure 1 (right).
 Thus the interior area limited by the bound curve is potentially excluded
 by the LEP-II experiment. 

 Following similar steps to the 
 single-photon case, we compute the limits from the single-Z channel 
\cite{ACDM}.
 In this case the total cross section is obtained by summing over 
 the three Z polarizations. The result for massless branons is now:
 $f \geq 52 n^{1/8} \mbox{GeV}$,
 and the exclusion plot can be found in Figure 1 (right).
 The excluded area limited by the bound curve is
 smaller than in the single photon analysis.
 This is an expected result since the Z
 coupling to
 the electron field is smaller than that of the photon,
 and the Z mass restricts the avalaible phase space.
 In fact, this restricts the search only to branons with
 masses:
 $M\leq(\sqrt{s}-M_Z)/2\simeq 57$ GeV. In the single-photon
 channel however, the kinematical range is larger
 $M\leq\sqrt{s}/2\simeq 103$ GeV.
 In  future colliders, working at very high
 centre-of-mass energy, the Z mass
 could be neglected $\sqrt{s}\gg M_Z$, and both processes will
 allow us to get similar bounds.
 \begin{figure}[h]  
\resizebox{14cm}{!}{\includegraphics{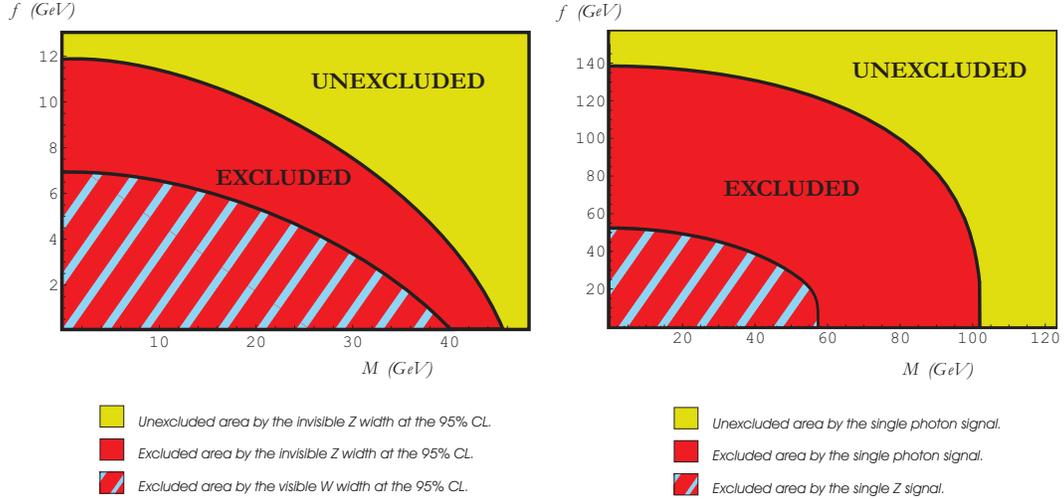}}
 \caption{\footnotesize{{\bf Left}: exclusion plot in the $f-M$ plane
 for single-branon models from LEP-I data.
 The dark area is excluded by Z
 invisible width data and the striped region is excluded by the
 visible $W^\pm$ decay.}
 \footnotesize{{\bf Right}: exclusion regions in the $f-M$ plane
 coming from single-photon events (dark region) and single-Z
 processes (striped area), assuming no observation at LEP-II.}}
 \end{figure}

 \section{Prospects for future linear colliders}

 Several proposals for the construction of  $e^+e^-$ linear
 colliders in the TeV range are currently under study. The TESLA
 (TeV Energy Superconducting Linear Accelerator), the NLC
 (Next Linear Collider)  and the JLC (Japanese Linear Collider)
 are examples of the first generation of these colliders, 
 whereas
 the CLIC (Compact Linear Collider)  would correspond to the
 second generation.  In this section we discuss the sensitivity of these
 colliders to a hypothetical branon signal. 

  The physics programme of the new linear collider projects includes
 the measurement of electroweak parameters with improved precision,
 such as the
 invisible Z width or the $W^\pm$ width.
 However, and since the deviations due to the
 presence of branons increase dramatically with energy, the
 largest sensitivity
 to a branon signal is expected in direct searches like
 single-$\gamma$ and
 single-Z.

  For this purpose the LEP-II study of the previous section is
 extended
 to higher center-of-mass energies. It is assumed that, at the time of
 construction of these accelerators, the theoretical and systematic
 uncertainties on Standard Model processes will be controlled at the level
 of the femtobarn. Under this assumption, we have estimated the 
 sensitivity limit by scaling the LEP-II estimate by the expected
 gain in statistics due to the luminosity improvement.

 The critical parameter in the analysis is the center-of-mass energy,
 $E_{CM}$. In the single photon channel, this leads to limits
 on $f$ for any branon mass $M<E_{CM}/2$.
 The results of a full study in the $(f,M)$ plane and in different
 experimental contexts are presented in Figure 2. 
For the single-Z, the bounds are less restrictive, and 
 the study is only applicable to branon masses below
 $(E_{CM}-M_Z)/2$, due
 to kinematic constraints (see also 
 Figure 2).
 \begin{center}
 \resizebox{14.5cm}{!}{\includegraphics{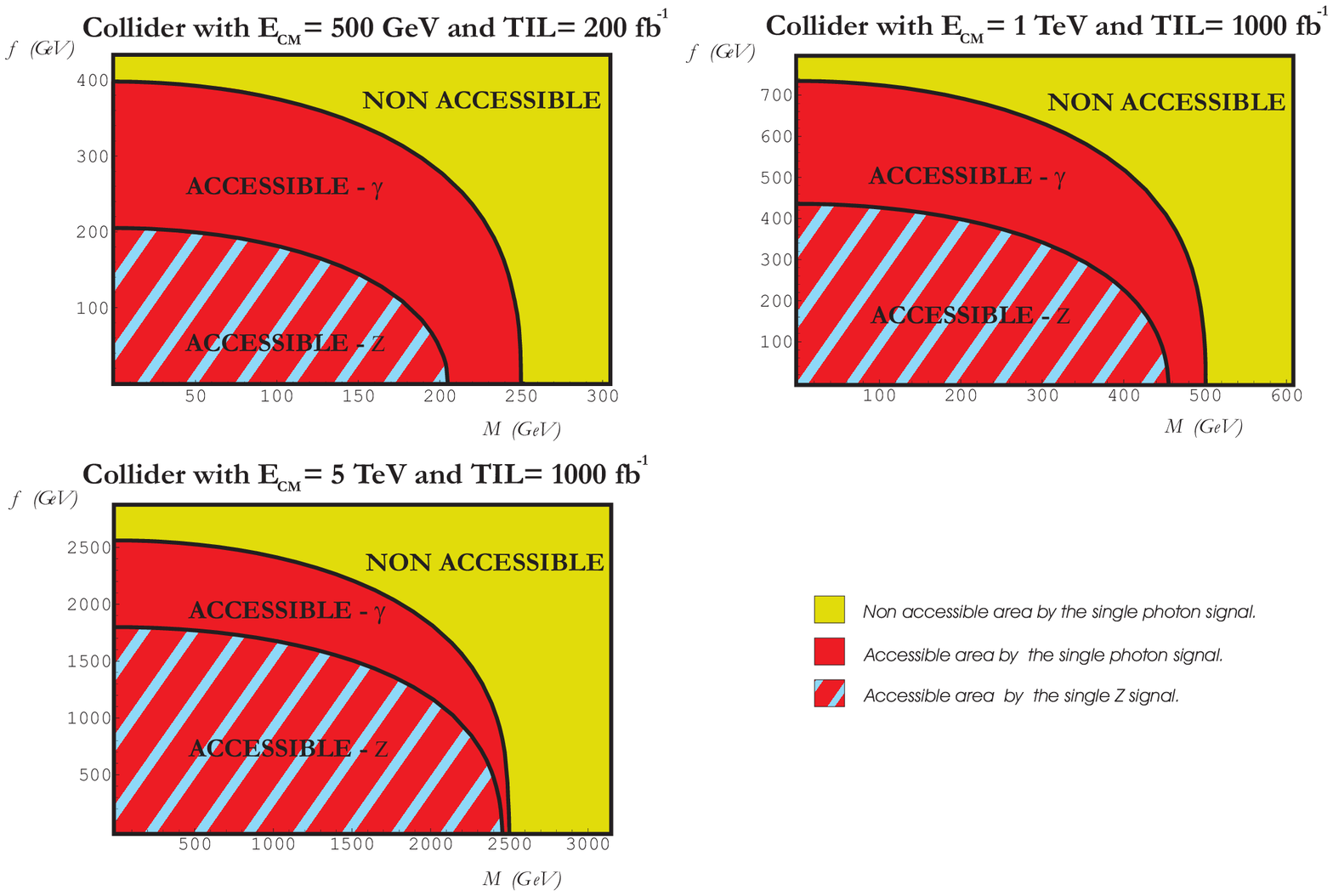}}
 \vspace*{-.2cm}
 \begin{figure}[h]  
 \caption{\footnotesize{
 Predicted experimental accessible
 regions in the $f-M$
 plane
 coming from single-photon events and single-Z
 processes, for future $e^+ e^-$ linear colliders with different
 center-of-mass energies
 ($E_{CM}$) and total integrated luminosities (TIL).}}
 \end{figure}
 \end{center}

 \section{Bounds from astrophysics and cosmology}

 Since branon effects grow strongly with energy, it
 is not surprising that the bounds from direct searches
that we have just obtained 
 with  ($\sqrt{s}\simeq 200$ GeV) are more constraining
 than the indirect ones, in which the energy scale
 is set by the Z mass ($M_Z\simeq 90$ GeV). In addition the 
 bounds presented here improve the astrophysical
 ones for massless branons,
 coming from energy loss in supernova 1987A, 
from which $f > 10$ GeV \cite{Kugo}.
For massive branons, the astrophysical limits are much less restrictive 
(see \cite{CDM}).

 Concerning cosmology, it
 is interesting to notice that
 the allowed range of parameters suggests that branons could have
 weak couplings and large masses. Since they are stable particles, this 
makes them natural dark matter
 candidates. In fact, an explicit calculation shows that
 their relic abundance can be cosmologically relevant and could
 account for the fraction of one third of the total energy density of
 the universe in the form of dark matter presently favoured by
 observations. 

 On the other hand, the present 
limits on dark matter abundance can be used
 also to restrict the parameters space of the theory. The results are 
 complementary to those of colliders. In the latter case, 
 as we have seen, the
 unexcluded region corresponds to small cross sections (large $f$) and/or 
 large branon masses, whereas having cosmological branon 
 abundances compatible with
 observations requires either large cross sections (small $f$) or small
 branon masses.
 These results will be presented elsewhere
 \cite{CDM}.

\section{Summary and conclusions}

Brane world scenarios are inspired on modern string (M) theory and offer new
insights on many fundamental problems in particle physics. If the
fundamental scale of gravity $M_D$ is of the order of 1 TeV, 
gravitons could
be produced in the next generation of colliders such as the LHC. Detailed
computations of the corresponding cross sections require to take into account
the recoil effects of the brane. In the flexible brane limit ($f\ll M_D$) the
only relevant modes are the SM particles and  the branons. Branon
production rates and their cosmological and astrophysical implications can be
determined in a model independent way in terms of the brane tension and the
branon mass.

\vspace{.5cm}

 {\bf Acknowledgements:} This work
 has been partially supported by the DGICYT (Spain) under the
 project numbers
 PB98-0782, AEN99-0305, FPA 2000-0956 and BFM2000-1326.
 A.D. acknowledges  the hospitality of the LBNL Theory
Group, where this work was partially done, economical support from the
Universidad Complutense del Amo Program and very especially the organizers of
the X Mexican School of Particles and Fields for their kind invitation 
and congratulates
Augusto Garc\'\i a and Arnulfo Zepeda for their 60th birthay.

\end{document}